\documentclass[a4paper,12pt,amssymb,amsmath,tightenlines]{article}
\usepackage{amsmath, amssymb, amsthm, latexsym, mathrsfs}
\usepackage{dsfont}
\usepackage{color}

\newcommand{\rd}{\mbox{$\rm d$}}

\newcommand{\e}{\textrm{e}}
\newcommand{\nn}{\nonumber}
\newcommand{\ind}{1{\hskip -2.5 pt}\hbox{I}}
\newtheorem{lem}{Lemma}[section]
\newtheorem{prop}{Proposition}[section]

\newtheorem{ass}{Assumption}[section]
\theoremstyle{definition}

\newtheorem{rem}{Remark}[section]
\numberwithin{equation}{section}
\renewcommand{\labelitemi}{$$}
\newenvironment{noindlist}
 {\begin{list}{\labelitemi}{\leftmargin=0em \itemindent=0em}}
 {\end{list}}

\title{\bf{On the Representation of \\ General Interest Rate Models as \\Square-Integrable Wiener Functionals}}
\begin{document}

\author{Lane P. Hughston and Francesco Mina}
\date{}
\maketitle
\begin{center}
Department of Mathematics, Imperial College London\\London SW7 2AZ, United Kingdom\\\end{center}

\begin{abstract}
In the setting proposed by Hughston \& Rafailidis (2005) we consider general interest rate models 
in the case of a Brownian market information filtration $(\mathcal{F}_t)_{t\geq0}$. Let $X$ be a square-integrable $\mathcal{F}_\infty$-measurable random variable, and assume the non-degeneracy condition  
that for all $t<\infty$ the random variable $X$ is not $\mathcal{F}_t$-measurable.
Let ${\sigma_t}$ denote the integrand appearing in the representation of $X$ as a stochastic integral, 
write $\pi_t$ for the conditional variance of $X$ at time $t$, and set $r_t = \sigma^2_t / \pi_t$. 
Then $\pi_t$ is a potential, and as such can act as a model for a pricing kernel (or state price density), 
where $r_t$ is the associated interest rate. Under the stated assumptions, we prove the following: (a) that 
the money market account process defined by $B_t = \exp (\int_0^t r_s \,ds)$ is finite almost surely at all 
finite times; and (b) that the product of the money-market account and the pricing kernel is a local martingale, and is a martingale provided a certain integrability condition is satisfied. 
The fact that a 
martingale is thus obtained shows that from any non-degenerate element of Wiener space satisfying the integrability condition we can construct an associated interest-rate model. The model thereby constructed is valid over 
an infinite time horizon, with strictly positive interest, and satisfies the relevant intertemporal relations 
associated with the absence of arbitrage. The results thus stated pave the way for the use of Wiener chaos 
methods in interest rate modelling, since any such square-integrable Wiener functional admits a chaos expansion, 
the individual terms of which can be regarded as parametric degrees of freedom in the associated interest rate 
model to be fixed by calibration to appropriately liquid sectors of the interest rate derivatives markets. 
\begin{center}
\end{center}
\end{abstract}
\vspace{.25cm}
\\ \vspace{.25cm}\\
Email: lane.hughston@imperial.ac.uk,~
francesco.mina08@imperial.ac.uk
%

\section{Introduction}
In the so-called ``chaotic approach" to interest rate modelling, Hughston \& Rafailides \cite{HUGHS} show that the general interest rate model, subject to a set of axioms, is fully characterized in the case of a Brownian filtration by the specification of a random variable $X \in L^2(\Omega, \mathcal{F}_{\infty}, \mathbb{P})$. Since such random variables admit a Wiener chaos expansion, the resulting interest rate models can be parametrized in a natural way by a collection of deterministic functions, thus leading to a rather general calibration methodology. In particular, once the chaos coefficients have been specified, the initial term structure, the volatility structure and the market price of risk are determined in the associated interest rate model. Further developments of the chaotic approach, with examples of various explicit models and calibration schemes, are reported by Brody \& Hughston \cite{BH3}, Rafailidis \cite {RAF}, Grasselli \& Hurd \cite{GRAS}, Tsujimoto \cite{TSU}, Grasselli \& Tsujimoto \cite{GRAS2}, and others. The present paper is concerned with the inverse problem: given an element of $L^2(\Omega, \mathcal{F}_{\infty}, \mathbb{P})$, are we able to construct an associated interest rate model?  

To begin, let us recall briefly the axiomatic scheme of Hughston \& Rafailidis \cite{HUGHS}. 
Let $(\Omega, \mathcal{F}, \mathbb{P})$ be a probability space equipped with the augmented filtration $(\mathcal{F}_t)_{t \geq 0}$ generated by a system of $n$ independent Brownian motions. All processes under consideration in what follows are assumed to be c\`adl\`ag. We introduce the following axioms:
\begin{noindlist}
\item{(A1)} There exists a non-dividend-paying money-market asset with price process $(B_t)_{t \geq 0}$ given by an expression of the form
\begin{equation} 
	B_t = \exp{\left(\int_0^t r_s\, \rd s \right)}\,,
\end{equation}
where the short-rate process $(r_t)_{t\geq 0}$ satisfies $r_t \geq 0$ almost surely for all $t \geq 0$.
\item{(A2)} There exists a process $(\pi_t)_{t\geq 0}$ satisfying $\pi_t >0$ almost surely for all $t\geq0$ such that for any asset with price process $(S_t)_{t\geq0}$ and cumulative dividend process $(\Delta_t)_{t\geq0}$, the associated ``deflated total value" process $(\bar{S}_t)_{t\geq0}$ defined by
\begin{equation} \label{axiom}
	\bar{S}_t = \pi_t S_t + \int_0^t \pi_s \,\rd \Delta_s
\end{equation}
is a martingale. The process $(\pi_t)$ is called the ``pricing kernel".
\item{(A3)} There exists an asset (a perpetual floating rate note) that offers a dividend rate that ensures that the value of the asset is constant.
\item{(A4)} A system of discount bonds $(P_{tT})$ exists for $0\leq t < \infty$ and $0\leq T < \infty$ with the property that for all $t \geq 0$ we have
\begin{equation}
	\lim_{T \to \infty} P_{tT} = 0\,, \qquad \mathbb{P}\text{-a.s.}\,.
\end{equation}
\end{noindlist}

Under these assumptions one can show that the pricing kernel admits a representation as a conditional variance. In particular we have the following, where we write $\mathbb{E}_t [\cdot]$ for conditional expectation with respect to $\mathcal{F}_t$. 
\begin{prop}
If axioms \rm{(A1)-(A4)} hold, then there exists a random variable $X\in L^2(\Omega, \mathcal{F}_{\infty}, \mathbb{P})$ such that
\begin{equation} \label{wiw}
	\pi_t = \mathbb{E}_t\left[(X - \mathbb{E}_t[X])^2\right].
\end{equation}
\end{prop}
\noindent
\textbf{Proof.} It follows from axioms (A1) and (A2) that the process $(\rho_t)_{t\geq0}$ defined by $\rho_t = \pi_t B_t$ is a martingale, and that $\rho_t >0$ almost surely for all $t \geq 0$. Thus we have $\pi_t = \rho_t/B_t$ and we see that $(\pi_t)$ is a supermartingale. Next, writing $(1_t )_{t \geq 0}$ for the value process of a unit floating rate note, we observe by use of (A2) and (A3) that the deflated total value process $(\bar 1_t)_{t\geq0}$ defined by
\begin{equation}
	\bar 1_t = \pi_t + \int_0^t \pi_sr_s \rd s
\end{equation}
is a martingale, from which we deduce that 
\begin{equation}
	\pi_t = \mathbb{E}_t[\pi_T] + \mathbb{E}_t\left[\int_t^T \pi_sr_s \rd s\right]\,.
\end{equation}
Since $\pi_tP_{tT} = \mathbb{E}_t[\pi_T]$ for $0\leq t < T$, it follows from (A4) that 
\begin{equation}
	\pi_t = \lim_{T \to \infty}\mathbb{E}_t\left[\int_t^T \pi_sr_s \rd s\right]\,,
\end{equation}
and hence 
\begin{equation}
	\pi_t = \mathbb{E}_t\left[\int_t^{\infty} \pi_sr_s \rd s\right]\,,
\end{equation}
by the conditional form of the monotone convergence theorem. Writing
\begin{equation}
	X = \int_0^{\infty} \sum_{\alpha= 1}^n \sigma^{\alpha}_t \,\rd W^{\alpha}_t\,,
\end{equation}
where $(W^{\alpha}_t)_{t\geq0}$ is the $n$-dimensional Brownian motion upon which the filtration is based, and where $(\sigma^{\alpha}_t)_{t\geq0}$ is any adapted vector-valued process satisfying
\begin{equation} \label{pain}
	\sum_{\alpha= 1}^n (\sigma^{\alpha}_t)^2 = \pi_tr_t \,,
\end{equation}
we thus obtain the desired relation (\ref{wiw}) by the use of the It\^{o} isometry.
\hfill$\Box$
\vspace{0.5cm}

The significance of the conditional variance representation (1.4) for the pricing kernel is that the resulting interest rate system is entirely determined by the random variable $X$. Thus, all of the information of the interest rate model is somehow ``compressed" into the specification of $X$. It is natural therefore to enquire about the extent to which the reverse construction holds. The purpose of this paper is thus to show in some detail how it is possible to construct an interest rate model from any $X$ in $L^2(\Omega, \mathcal{F}_{\infty}, \mathbb{P})$ satisfying certain stated conditions. 
The structure of the paper is as follows. This introductory section concludes with a series of remarks commenting on various aspects of the Hughston-Rafailidis scheme, with emphasis on the cash flows offered by the financial instruments under consideration, both in the discrete case and the continuous case. The geometric Brownian motion model is studied in some detail as an example. In Section 2 we consider the inverse problem. A non-degeneracy condition is imposed in Assumption 2.1 to ensure that the interest rate model resulting from the chosen square-integrable Wiener functional extends to an infinite time horizon. With this assumption, and by use of the technical Lemma \ref{lemma}, it is shown in Proposition \ref{twopointone} that the conditional variance of any non-degenerate element of $ L^2(\Omega, \mathcal{F}_\infty, \mathbb{P})$ defines a strictly positive type-D potential, and hence determines a pricing kernel $(\pi_t)$. The resulting pricing kernel is used to construct a discount bond system $(P_{tT})$, a set of instantaneous forward rates $(f_{tT})$, a short rate process $(r_{t})$, a floating rate note of constant unit value $(1_t)$, a  natural numeraire process $(\xi_{t})$, a money market account process $(B_t)$, and a family of options on discount bonds of various strikes and maturities. The discount bonds, the floating rate note, the natural numerarie, and the options are shown to satisfy (A2). In Section 3 we show in Proposition \ref{threepointone} that the money market account is finite almost surely, and in Proposition \ref{threepointtwo} we prove that the product of the pricing kernel and the money market account is a local martingale.  In Proposition \ref{3.3} we establish (by use of the technical Lemma \ref{quotient}) an integrability condition that is sufficient to ensure that the money market account satisfies the intertemporal conditions, and we conclude with Remark \ref{financialinterp}, which supplies a financial interpretation to the integrability condition. 

\begin{rem} \label{ciao}
In the Hughston-Rafailidis scheme the market is represented by a collection of financial assets, each of which is characterised by the specification of a value (or price) process $(S_t)_{t\geq0}$ and a cumulative cash-flow (or ``dividend") process $(\Delta_t)_{t\geq 0}$, both taking values in $\mathbb{R}$. In the present paper we generalize axiom (A2) of \cite{HUGHS} to allow for the inclusion of discrete dividends. In general, the assets under consideration need not have the ``limited liability" property, and one can think of the pair $(S_t, \Delta_t)_{t\geq 0}$ as representing a position in a financial contract involving both positive and negative cash flows. As a simple example of a contract involving both positive and negative (net) cash flows, consider a position in a standard interest rate swap, where an agent receives a fixed rate and pays a floating rate. The value of such a position at any given time can be positive or negative. On the other hand, in the situation of a limited liability asset we have $S_t \geq 0$ for all $t\geq 0$ and $(\Delta_t)_{t \geq 0}$ is an increasing process . The assumed c\`adl\`ag property embodies the idea that the price of such an asset goes ``ex-dividend" the instant that the dividend is paid. 
\end{rem}
\begin{rem} \label{taos}
As an illustration of the set-up described in Remark 1.1, consider, for example,  the case of a limited liability asset that pays a single random cash flow $H_T \geq 0$ at a fixed time $T$, and drops to the value zero when the payment is made. Write $(S_t)_{t\geq0}$ for the price process and $(\Delta_t)_{t\geq0}$ for the cumulative dividend process. We require that $H_T$ should be $\mathcal{F}_T$-measurable and that the deflated total value process $\bar{S}_t$ defined by $\bar{S}_t = \pi_t S_t$ for $t < T$ and $\bar{S}_t = \pi_T H_T$ for $t \geq T$ should be a martingale. A calculation using the martingale condition then shows that the price process is given by  
\begin{equation}
S_t =  \ind ( t < T) \frac{1}{\pi_t} \mathbb{E}_t [\pi_T H_T]\, .
\end{equation}
The associated cumulative dividend process is given by 
\begin{equation}
\Delta_t = \ind (t \geq T) H_T \,.
\end{equation}
We note that the deflated total value process is a uniformly-integrable martingale:
\begin{equation}
\bar{S}_t = \mathbb{E}_t [\pi_T H_T] \,.
\end{equation}
In what follows we refer to the various conditions that arise as a consequence of (A2) as ``intertemporal relations".
\end{rem}

\begin{rem}
By a unit $T$-maturity discount bond, we mean a financial instrument of the type described in Remark 1.2, with $H_T = 1$. Writing as $P_{tT}$ for the price at time $t$ of a $T$-maturity discount bond, it follows as a consequence of (A2) that $\lim_{t \to T} P_{tT} = 1$, and that $P_{tT}=0$ for all $t\geq T$. The resulting discount bond price system 
$(P_{tT})$ is thus defined for all $t\geq0$ and all $T\geq0$. Our convention differs slightly from the more usual one (in which a $T$-maturity discount bond matures to take the value unity at $T$), but is perhaps better because it allows for a consistent and transparent treatment of the relevant cash flows. 
\end{rem}

\begin{rem}
In the case of the standard geometric Brownian motion (GBM) model for asset pricing used as the basis of the Black-Scholes theory, the pricing kernel takes the form 
\begin{equation} \label{killingmemore}
	\pi_t = \e^{-rt - \lambda W_t - \frac{1}{2}\lambda^2 t}\,,
\end{equation}
where $r$ and $\lambda$ are constants. The price of a typical limited-liability asset with constant volatility and paying a continuous proportional dividend at a constant rate $\delta$ is given by 
\begin{equation}
S_t = S_0 \, \e^{(r-\delta + \lambda \sigma)t + \sigma W_t - \frac{1}{2}\sigma^2 t }\, ,
\end{equation} 
and the associated cumulative dividend process is given by 
\begin{equation}
\Delta_t = \delta \int_0^t S_u\, \rd u\,.
\end{equation}
If $\delta = 0$, it is straightforward to check that $(\pi_t S_t)_{t \geq 0}$ is a martingale. If $\delta \neq 0$ we obtain
\begin{equation} 
\bar{S}_t  = S_0 \, \e^{-\delta t + (\sigma - \lambda)W_t  -\frac{1}{2}(\sigma - \lambda)^2 t} + \delta S_0 \int_0^t \e^{-\delta u + (\sigma - \lambda)W_u - \frac{1}{2}(\sigma - \lambda)^2u} \rd u
\end{equation}
where  $\bar{S}_t $ is defined as in equation (\ref{axiom}), and a calculation allows us to deduce that
\begin{equation} \label{8thJune}
\bar{S}_t  = S_0 \left[ 1 + (\sigma - \lambda) \int_0^t \e^{-\delta u + (\sigma - \lambda)W_u - \frac{1}{2}(\sigma - \lambda)^2 u} \, \rd W_u\right]\,.
\end{equation}
In particular, an application of Fubini's theorem shows that 
\begin{equation}
\mathbb{E}\left[\int_0^t (\e^{-\delta u + (\sigma - \lambda)W_u - \frac{1}{2}(\sigma - \lambda)^2 u})^2 \, \rd u\right] < \infty
\end{equation}
for all $t < \infty$, and hence that the stochastic integral on the right-hand side of (\ref{8thJune}) exists and defines a martingale. Thus we see that the conditions of (A2) are satisfied in the GBM model for the given processes $(\pi_t)$, $(S_t)$, and $(\Delta_t)$. It is interesting to note that a necessary and sufficient condition for $(\bar{S}_t)$ to be a uniformly integrable martingale is $\delta > \frac{1}{2} (\sigma - \lambda)^2$.
\end{rem}

\begin{rem}
The pricing kernel of the geometric Brownian motion model arises if we take the random variable $X$ to be of the form
\begin{equation} \label{killingme}
	X = \sqrt{r}\int_0^{\infty} \e^{-\frac{1}{2}rt - \frac{1}{2}\lambda W_t - \frac{1}{4}\lambda^2 t}\, \rd W_t\,.
\end{equation}
Indeed, it is a simple exercise to check that 
\begin{equation}
\mathbb{E}\left[\int_0^{\infty} \e^{-rt - \lambda W_t - \frac{1}{2}\lambda^2 t}\, \rd t \right]  < \infty\, ,
\end{equation}
and hence (i) that the stochastic integral on the right-hand side of equation (\ref{killingme}) exists, and (ii) that $\mathbb{E}[X^2] < \infty$. A calculation shows that the conditional variance of $X$ is given by (\ref{killingmemore}), and a glance at (\ref{killingme}) allows one to check that ${\sigma_t}^2/ \pi_t = r$, and hence that $B_t = \e^{rt}$. It follows immediately from (\ref{killingmemore}) that $(\pi_t B_t)_{t \geq 0}$ is a martingale. Likewise for fixed $T$ we deduce that $P_{tT} = \e^{-(T-t)r}$ for $0\leq t < T$, and hence that the deflated total value process
 \begin{equation}
 \bar{P}_{tT} = \ind(t < T)\pi_t P_{tT} + \ind(t \geq T)\pi_T\,,
 \end{equation}
 defined for $t\geq 0$, is a martingale. Thus we see that the money market account and the discount bond system both satisfy the conditions of (A2) in the GBM model. 
 \end{rem}
\begin{rem} 
As a candidate for a floating rate note, in the geometric Brownian motion model, we consider an instrument with a principal of unity that pays a continuous dividend at the rate $r$. We need to check that the deflated total value process, given by  $\bar 1_t = \pi_t + r \int_0^t \pi_s \rd s$, $t\geq0$, is a martingale. A calculation shows that
\begin{eqnarray}
\pi_t + r\int_0^t \pi_s \rd s &=& \e^{-rt - \lambda W_t - \frac{1}{2}\lambda^2t} + r \int_0^t \e^{-ru - \lambda W_u - \frac{1}{2}\lambda^2u}\rd u \\
&=& 1 - \lambda \int_0^t \e^{-ru -\lambda W_u - \frac{1}{2}\lambda^2 u}\, \rd W_u\,.
\end{eqnarray}
In particular, one can verify that 
\begin{equation}
\mathbb{E}\left[\int_0^t (\e^{-ru -\lambda W_u - \frac{1}{2}\lambda^2 u})^2\, \rd u\right] < \infty
\end{equation}
for all $t < \infty$, and hence that (i) the stochastic integral on the right-hand side of equation (1.24) exists, (ii)  a martingale is obtained, and (iii) the conditions of (A2) are met for the floating rate note in the GBM model.
\end{rem}

\begin{rem}
It should be evident from equation (\ref{pain}) that the vector-valued process $(\sigma^{\alpha}_t)$ appearing as the integrand in the expression for $X$ is uniquely determined up to an adapted rotation of the form 
\begin{equation}
	\sigma^{\alpha}_t \mapsto \sum_{\beta = 1}^n L_{\beta t}^{\alpha} \sigma^{\beta}_t
\end{equation}  
where
\begin{equation}
\sum_{\alpha = 1}^n \sum_{\alpha^{\prime} = 1}^n \delta_{\alpha\alpha^{\prime}}L_{\beta t}^{\alpha}L_{\beta^{\prime}t}^{\alpha^{\prime}} = \delta_{\beta \beta^{\prime}}\,.
\end{equation}
We observe that the matrix process $(L_{\beta t}^{\alpha})_{t \geq 0}$ need not be continuous. In particular, in the case of a one-dimensional Brownian motion, the process $(\sigma_t)$ is unique up to a transformation of the form $\sigma_t \mapsto u_t \sigma_t$, where $(u_t)_{t \geq 0}$ is an adapted ``unit" process, whose state space is $\{1, -1\}$.
\end{rem}

\section{Construction of an interest-rate market}
We consider a probability space $(\Omega, \mathcal{F}, \mathbb{P})$ with the standard augmented filtration $(\mathcal{F}_t)_{t \geq 0}$ generated by a Brownian motion $(W_t)_{t \geq 0}$, which for simplicity in the discussion that follows we henceforth take to be one-dimensional (generalization to the $n$-dimensional case is straightforward). Let $X$ be a real-valued square-integrable $\mathcal{F}_\infty$-measurable random variable. 
\begin{ass}
The random variable $X \in L^2(\Omega, \mathcal{F}_\infty, \mathbb{P})$ is assumed to be non-degenerate in the sense that $X$ is not $\mathcal{F}_t$-measurable for $0\leq t < \infty$. 
\end{ass}

\noindent Thus $X$ is a function of the entire path of the Brownian motion. We shall show how each such choice of $X$ determines a model for a ``market" for various interest-rate related instruments. In particular, from $X$ we shall construct a pricing kernel, a family of discount bonds of all maturities, a floating rate note, a money market account, a natural numeraire, and a variety of other interest-rate related instruments, and determine conditions under which (A2) holds. The construction proceeds as follows. 

Since $X \in L^2(\Omega, \mathcal{F}_\infty, \mathbb{P})$, it is a standard result (see, e.g., Davis 2005, Oksendal 2010, Revuz \& Yor 2001) that $X$ can be put in the form    
\begin{equation} \label{intrep1}
	X = X_0 + \int_0^\infty \sigma_s \, \rd W_s,
\end{equation}
where $X_0 = \mathbb{E}[X]$. Here the integrand $(\sigma_t)_{t \geq 0}$ is adapted to $(\mathcal{F}_t)$ and satisfies 
\begin{equation} \label{intrep}
	\mathbb{E}\left[\int_0^\infty \sigma_s^2 \, \rd s \right] < \infty .
\end{equation}
Then the process $(X_t)_{t \geq 0}$ defined by the conditional expectation $X_t = \mathbb{E}_t[X]$ is a martingale, and we have
\begin{equation} \label{intrep2}
	X_t = X_0 + \int_0^t \sigma_s \, \rd W_s\,.
\end{equation}
To construct the pricing kernel, we let $(\pi_t)_{t\geq0}$ be given by the conditional variance. Thus $\pi_t$ is taken to be of the form
\begin{equation} \label{pk}
	\pi_t = \mathbb{E}_t\left[(X - \mathbb{E}_t[X])^2\right].
\end{equation}
We recall (Meyer 1966) that by a potential we mean a non-negative right-continuous supermartingale satisfying
\begin{equation}
	\lim_{t \to \infty} \mathbb{E}[\pi_t] = 0 \,.
\end{equation}
Thus a right-continuous supermartingale $(\pi_t)$ is a potential if and only if $\pi_t \geq 0$ for $0 \leq t < \infty$, and  $(\pi_t)$ converges to $0$ both almost surely and in the $L^1$ norm. By a type-D potential, we mean a potential that takes the form 
\begin{equation}
	\pi_t = \mathbb{E}_t[A_{\infty}] - A_t \,,
\end{equation} 
where $(A_t)_{t\geq0}$ is an integrable increasing process. By an increasing process we mean a real-valued adapted process $(A_t)_{t\geq0}$, with non-decreasing right-continuous paths, such that $A_0 = 0$ almost surely, and such that the random variables $A_t$ are integrable. An increasing process $(A_t)_{t\geq0}$ is then said to be integrable if $\sup_t  \mathbb{E}[A_t]$ is finite, or, equivalently, if  $\mathbb{E}[A_\infty] < \infty$, where $A_\infty= \lim_{t \to \infty} A_t$.
\noindent
\begin{prop} \label{twopointone}
The process $(\pi_t)_{t\geq0}$ is a strictly-positive type-D potential.
\end{prop}

\noindent
\textbf{Proof.} To see that the process defined by (\ref{pk}) is a supermartingale we observe that 
\begin{equation} \label{cts}
	\pi_t = \mathbb{E}_t[X^2] - \left(\mathbb{E}_t[X] \right)^2.
\end{equation} 
Since $\{(\mathbb{E}_t[X])^2\}$ is a submartingale, it follows that $(\pi_t)$ is the sum of a martingale and a supermartingale, and is thus itself a supermartingale. Next, we note that since $\{\mathbb{E}_t[X]\}$ and $\{\mathbb{E}_t[X^2]\}$ are uniformly integrable martingales, they converge respectively to $X$ and $X^2$ almost surely, and it follows that $(\pi_t)$ converges to zero almost surely. 
When $X$ is of the form (\ref{intrep1}) and $(\mathcal{F}_t)$ is the Brownian filtration, then a short calculation making use of the It\^{o} isometry shows that 
\begin{equation}
	\pi_t = \mathbb{E}_t \left[\int_0^{\infty}  \sigma_s^2 \, \rd s \right] - \int_0^{t}  \sigma_s^2 \, \rd s\,, 
\end{equation}
and it follows by equation (2.2) that $(\pi_t)$ is a type-D potential, as claimed. The fact that $\pi_t > 0$ almost surely for $0 \leq t < \infty$ is a consequence of Assumption (2.1). For suppose it were the case that $\pi_t = 0$ for some value of $t < \infty$. Then for that value of $t$ we would have $\mathbb{E}_t\left[(X - \mathbb{E}_t[X])^2\right] =0$ almost surely, which would imply $\mathbb{E}\left[(X - \mathbb{E}_t[X])^2\right] =0$, and therefore $X = \mathbb{E}_t[X]$ almost surely; and thus $X$ would be $\mathcal{F}_t$-measureable, which would contradict Assumption (2.1). 
\hfill$\Box$

\vspace{0.5cm}
To see that $X = \mathbb{E}_t[X]$ almost surely implies that $X$ is $\mathcal{F}_t$-measureable in the proof above, one observes the following.
\noindent
\begin{lem} \label{lemma}
Let $X$ and $Y$ be real-valued random variables on $(\Omega, \mathcal{F}, \mathbb{P})$, and suppose that $Y \in m\mathcal{F}$ and $X= Y$ almost surely. Then $X\in m\mathcal{F}^{\mathbb{P}}$ where $\mathcal{F}^{\mathbb{P}}$ is the completion of $\mathcal{F}$ with respect to $\mathbb{P}$. In particular, if $\mathcal{F} = \mathcal{F}^{\mathbb{P}}$, then the conditions $X= Y$ almost surely and $Y\in m\mathcal{F}$ together imply $X \in m\mathcal{F}$.
\end{lem}
\noindent
\textbf{Proof.} 
A useful characterization of $\mathcal{F}^{\mathbb{P}}$ is given by $\mathcal{F}^{\mathbb{P}} = \mathcal{H}$, where
\begin{equation}
	\mathcal{H} = \{F \subset \Omega \mid \exists\, G \in \mathcal{F} \text{ such that } F \triangle G \in \mathcal{N}^{\mathbb{P}}\}\, ,
\end{equation}
$\mathcal{N}^{\mathbb{P}}$ represents the $\mathbb{P}$-null sets, and $\triangle$ denotes the symmetric difference. For a proof of this fact we refer to \cite{KS2}.
To prove Lemma \ref{lemma},  we need to show that $X^{-1}(B) \in \mathcal{H}$ for all $B \in \mathcal{B}(\mathbb{R})$. We observe that $X^{-1}(B) \triangle Y^{-1}(B) \in \mathcal{N}^{\mathbb{P}}$ since $X^{-1}(B) \triangle Y^{-1}(B) \subset \{X \neq Y\}$ and $\mathbb{P}(X \neq Y) = 0$. The result follows since $Y^{-1}(B) \in \mathcal{F}$. 
\hfill$\Box$
\vspace{0.5cm}

Thus starting with a non-degenerate $\mathcal{F}_{\infty}$-measurable element of $L^2(\Omega, \mathcal{F}, \mathbb{P})$ we obtain a process $(\pi_t)$ which we identify as the pricing kernel. Given $(\pi_t)$, one can proceed to construct a number of interest-rate related financial instruments. First, we construct the associated discount bond system $(P_{tT})$, where $P_{tT}$ denotes the random value at time $t$ of a discount bond that matures at time $T$ to deliver one unit of currency.
By Remark \ref{taos}, for each fixed maturity date $T$, the price process of a $T$-maturity discount bond is given for $ t \geq 0 $ by
\begin{equation}
	P_{tT} = \ind(t < T) \frac{1}{\pi_t}\mathbb{E}_t[\pi_T],
\end{equation}
and for the associated cumulative dividend rate process we have $\Delta_t = \ind(t \geq T)$. For each maturity date $T$ the deflated total value process is a uniformly integrable martingale, given by $\bar P_{tT} = \mathbb{E}_t[\pi_T]$, and it follows that the discount bond system thus constructed satisfies the conditions of (A2). 

The existence of the discount bond system is guaranteed because $\pi_t >0$ and $\pi_t \in L^1(\Omega, \mathcal{F}, \mathbb{P})$ for all $t \geq 0$. More specifically, making use of the relation 
\begin{equation} \label{check}
	\pi_t = \mathbb{E}_t \left[ \int_t^\infty \sigma_s^2\, \rd s \right]\, ,
\end{equation}
we obtain the following expression for the discount bond prices: 
\begin{equation} \label{bond prices}
	P_{tT}= \ind(t<T)\frac{\mathbb{E}_t\left[\int_T^{\infty} \sigma_s^2\, \rd s\right]}{\mathbb{E}_t\left[\int_t^\infty \sigma_s^2\, \rd s\right]}\, ,
\end{equation}
and hence 
\begin{equation} \label{bond prices 2}
P_{tT}= \ind(t<T)\frac{\int_T^{\infty} \mathbb{E}_t[\sigma_s^2]\, \rd s}{\int_t^{\infty} \mathbb{E}_t[\sigma_s^2]\, \rd s}\, .
\end{equation}
This is the so-called Flesaker-Hughston representation of the discount bond system; see Bjork \cite{BJORK}, Cairns \cite{CAIRNS}, Flesaker \& Hughston \cite{FH1}, Rutkowski \cite{RUT}, Hunt \& Kennedy \cite{HK}, Jin \& Glasserman \cite{JG}, Musiela \& Rutkowski \cite{MR}, Rogers \cite{ROG}.
Given the discount bond system, we can proceed to introduce various associated interest rates.
We recall that the instantaneous forward rate $f_{tT}$ is defined for $0\leq t < T < \infty$ by
\begin{equation}
	f_{tT} = - \frac{\partial}{\partial T} \ln P_{tT}\,,
\end{equation}
provided that the member on the right exists. In the present context an application of Fubini's theorem allows us to interchange the conditional expectation and the integral in the numerator of the right-hand side of equation (\ref{bond prices}), thus (a) leading to (\ref{bond prices 2}) and (b) ensuring that $f_{tT}$ exists and is given by the following expression:
\begin{equation} \label{ciao}
	f_{tT} = \frac{\mathbb{E}_t[\sigma_T^2]}{\mathbb{E}_t\left[\int_T^{\infty} \sigma_s^2\, \rd s\right]}\,.
\end{equation}
The associated short rate process $(r_t)_{t \geq 0}$ is then defined by $r_t = \lim_{t \to T} f_{tT}$, and it follows from equation $(\ref{ciao})$ that
\begin{equation}
	r_t = \frac{\sigma_t^2}{\mathbb{E}_t\left[\int_t^\infty \sigma_s^2\, \rd s\right]}\, ,
\end{equation}
or equivalently 
\begin{equation}
	r_t = \frac{\sigma_t^2}{\pi_t}\,,
\end{equation}
a formula that we shall use in what follows.

Given the interest rate $(r_t)$, one can proceed to introduce two further assets: (a) the floating rate note, and (b) the money market account. The latter of these requires rather careful consideration, and will be treated in the next section. We consider here the case of a floating rate note that pays a continuous dividend and maintains a constant value of unity. 

As in the previous section, let  us write $(1_t)_{t \geq 0}$ for the value process of a unit floating rate note. Elementary arguments (see, e.g., Hughston \& Rafailides 2005) show that the dividend rate offered by such an instrument must be the short rate. Our goal is to show that the conditions of (A2) are satisfied. Thus we need to check that the  associated deflated total value process  $(\bar 1_t)_{t \geq 0}$ defined by 
equation (1.5)
is a martingale. But it follows immediately from equations (1.5) and (\ref{check}) that
\begin{equation} \label{lasteqn}
	\bar 1_t =  \mathbb{E}_t\left[ \int_0^\infty \sigma_s^2\, \rd s\right]\,,
\end{equation}
which shows that $(\bar 1_t)$ is a uniformly integrable martingale, and hence that the intertemporal conditions of (A2) are satisfied.

We note, incidentally, that the role of ``cash" in economic thinking has always been somewhat contentious. Nevertheless one point of view which certainly has at least some validity is to think of ordinary money as being a kind of floating rate note. It has a constant value (in nominal terms), and pays a continuous dividend in the form of the convenience and liquidity we take for granted when we carry it.  

Continuing with our discussion of various assets associated with the interest rate market, next we introduce the so-called ``natural numeraire" (or benchmark portfolio) which is a non-dividend-paying asset with price process $(\xi_t)_{t \geq 0}$ given by $\xi_t = 1/\pi_t$. It is evident that the natural numeraire asset satisfies the intertemporal relations since the product $(\xi_t\pi_t)$ is a martingale. The natural numeraire has the following property. Let $(S_t)$ and $(\Delta_t)$ denote the price process and cumulative dividend process of any asset satisfying (A2). Then the associated deflated total value process $(\bar{S}_t)$, which is a martingale, given by 
\begin{equation}
	\bar{S}_t = \frac{S_t}{\xi_t} + \int_0^t \frac{1}{\xi_s}\, \rd \Delta_s
\end{equation}
can at each time $t$ be expressed as the sum of: (i) the value of the asset at that time, expressed in units of the natural numeraire, and (ii) the dividends paid to date, each expressed in units of the value of the natural numeraire at the time the dividend is paid.  

It should be apparent  that a variety of derivatives can be introduced based on the instruments already constructed. As an example, we consider a call option on a discount bond. Let the bond maturity be $T$ and the option maturity $t$, with $t < T$. Then the payout of an option with strike $K$ is $H_t = (P_{tT} - K)^{+}$, where $0< K < 1$. Clearly $\mathbb{E}[\pi_t H_t] < \infty$, since $0\leq P_{tT} < 1$. Therefore Remark \ref{taos} applies, and if we write $(C_s)_{0\leq s < t}$ for the value process of the option, then we have
\begin{equation}
C_s = \ind(s<t)\frac{1}{\pi_s}\mathbb{E}_s[\pi_t (P_{tT} -K)^{+}].
\end{equation}
The associated cumulative dividend process is given by 
\begin{equation}
\Delta_s = \ind(s\geq t)(P_{tT} -K)^{+}\,, 
\end{equation}
and a calculation shows that the deflated total value process $(\bar C_s)_{s \geq 0}$ is a uniformly integrable martingale, given by $\bar C_s = \mathbb{E}_s[\pi_t (P_{tT} -K)^{+}]$, which confirms that (A2) is satisfied. 

\section{Existence of the Money Market Asset}
We turn now to the construction of the money market account. The goal is twofold. First we need to show that the integral $\int_0^t r_s\, \rd s$ is finite almost surely for all $t\geq 0$. That this is the case may not be obvious since $r_t = \sigma_t^2 / \pi_t $, and $\pi_t$ is converging to zero. Then we need to find conditions that are sufficient to ensure that the unit-initialized money market process $(B_t)_{t \geq0}$ satisfies the intertemporal relations of (A2). We proceed as follows.
Let the money-market account process be defined by 
\begin{equation} \label{mma}
	B_t = \exp{\left(\int_0^t r_s\, \rd s \right)}\,,
\end{equation}
where $(r_t)$ is given by $r_t = \sigma_t^2/ \pi_t$, where $(\sigma_t)_{t\geq 0}$ is an adapted process satisfying equation (\ref{intrep}), 
and where $(\pi_t)$ is given by equation (\ref{check}). 
\begin{prop} \label{threepointone}
$\mathbb{P}(B_t < \infty) = 1$ for all $t \geq 0$.
\end{prop}  
\noindent
\textbf{Proof.}
We observe that the short rate can be expressed in the form  $r_t = \sigma_t^2 \xi_t$, for $t \geq 0$, where $\xi_t$ is the natural numeraire. By writing $\pi_t$ in the form $(\ref{cts})$ and making use of the martingale representation theorem we deduce that $\pi_t$ is continuous as a function of $t$. Thus $\xi_t$ is also continuous. Hence, for any fixed $t$ there exists a set $\tilde{\Omega} \subset \Omega$ with $\mathbb{P}(\tilde{\Omega} ) =1$ such that for all $\omega \in \tilde{\Omega}$  we can find a real $M = M(\omega, t)$ satisfying $\xi_s(\omega) \leq M$ for all $s \in [0,t]$, since any continuous function is bounded on a compact set. It follows from (\ref{intrep}) that 
\begin{equation}
	\mathbb{P}\left(\int_0^t \sigma_s^2\, \rd s < \infty \right) = 1\,,
\end{equation}
and hence we can find an $\bar{\Omega} \subset \Omega$ with probability one such that for all $\omega \in \bar{\Omega}$ we have 
\[
	\int_0^t \sigma_s^2 (\omega)\, \rd s < \infty.
\]
Clearly $\mathbb{P}(\tilde{\Omega} \cap \bar{\Omega}) = 1$. Let us therefore take $\omega \in \tilde{\Omega} \cap \bar{\Omega}$. By doing so we obtain
\begin{equation}
	\int_0^t r_s (\omega)\, \rd s = \int_0^t \sigma_s^2 (\omega)\xi_s (\omega)\, \rd s \leq M \int_0^t \sigma _s^2 (\omega)\, \rd s < \infty.
\end{equation}
Therefore $\mathbb{P}(B_t < \infty) = 1$ for all $t \geq 0$.
\hfill$\Box$
\vspace{0.5cm}

We have thus shown that the money market account is well-defined. It remains to be determined whether the conditions of (A2) are satisfied. Our strategy will be as follows. In Proposition 3.2 we establish that the product of the money market account with the pricing kernel is a local martingale. Then in Proposition \ref{3.3} we determine a condition under which this local martingale is a martingale.
\begin{prop} \label{threepointtwo}
Let the money market account value process $(B_t)$ be defined by equation $(\ref{mma})$, where $r_t = \sigma_t^2/\pi_t$ and $\pi_t = \mathbb{E}_t\left[\int_t^\infty \sigma_s^2\, \rd s\right]$. Then the process $(\rho_t)_{t \geq 0}$ defined by $\rho_t = \pi_tB_t $ is a local martingale. 
\end{prop}

\noindent
\textbf{Proof.} We recall that the pricing kernel can be written in the form 
\begin{equation}
	\pi_t = \bar 1_t - \int_0^t \sigma_s^2 \, \rd s\,
\end{equation}
where $(\bar 1_t)_{t\geq 0}$ is the uniformly integrable martingale defined by equation (\ref{lasteqn}).
By the martingale representation theorem we know that there exists an adapted process $(\theta_s)_{s \geq0}$ such that  
\begin{equation}
	\bar 1_t = \bar 1_0 + \int_0^t  \theta_s \, \rd W_s\, ,
\end{equation}
and such that  $(\theta_s)$ satisfies 
\begin{equation}
	\mathbb{P}\left(\int_0^\infty \theta_s^2 \, \rd s < \infty\right) = 1\,.
\end{equation}
Thus, setting $\lambda_t = -\theta_t/ \pi_t$, after a short calculation we deduce that  
\begin{equation}
	\rd \pi_t = - r_t \pi_t \rd t - \lambda_t \pi_t \rd W_t\,.
\end{equation}
Now let us consider the product 
\begin{equation} \label{three stars}
\rho_t = \pi_tB_t \,.
\end{equation} 
If 
\begin{equation} \label{star}
	\mathbb{P}\left(\int_0^t \lambda_s^2 \rho_s^2 \, \rd s < \infty\right) = 1,
\end{equation}
then it is an exercise in stochastic calculus to show that
\begin{equation}
	\rho_t = \rho_0 - \int_0^t \lambda_s \rho_s\, \rd W_s\, ,
\end{equation}
and hence that $(\rho_t)$ is a local martingale. To see that (\ref{star}) holds we proceed as follows. First we observe that $\lambda_t^2 \rho_t^2 = \theta_t^2 B_t^2$. For any choice of $\omega \in \Omega$, the function $\{ s \mapsto B_s(\omega) \}$ is continuous and hence, for fixed $t \geq 0$, there exists a real $M= M(\omega, t)$, such that $B_s(\omega) < M$ for all $s$ in the interval $[0,t]$.
We introduce the set 
\begin{equation}
{\Omega^\dagger} := \left\{\int_0^t \theta_s^2(\omega)\, \rd s < \infty\right\}\,, 
\end{equation}
which has probability one. Then for all $\omega \in{\Omega^\dagger}$ we have
\begin{equation}
	\int_0^t \theta_s^2 (\omega)B_s^2(\omega)\, \rd s < M^2\int_0^t \theta_s^2(\omega)\, \rd s < \infty\,,
\end{equation}
and as a consequence we see that (\ref{star}) holds, and thus that $(\rho_t)$ is a local martingale.
\hfill$\Box$
\vspace{0.5cm}

In the consideration of the intertemporal relations for the money market account, to which we now turn, we require the following Lemma.
\begin{lem} \label{quotient}
Let $X$ be a positive, integrable, random variable on a probability space $(\Omega, \mathcal{F}, \mathbb{P})$, and let $\mathcal{G}$ be a sub-$\sigma$-algebra of $\mathcal{F}$. Then  the random variable $X/\mathbb{E}[X|\mathcal{G}]$ is integrable and 
\begin{equation}
	\mathbb{E}\left[\frac{X}{\mathbb{E}[X|\mathcal{G}]}\right] = 1\,.
\end{equation}
\end{lem}
\noindent
\textbf{Proof.} We start by observing that 
\begin{equation}
	\lim_{n \to \infty} \left(\frac{1}{\mathbb{E}[X|\mathcal{G}]} \wedge n\right) \, X = \frac{X}{\mathbb{E}[X|\mathcal{G}]} \qquad \text{a.s.} .
\end{equation}
Hence, an application of the monotone convergence theorem shows that
\begin{equation}
	\mathbb{E}\left[\frac{X}{\mathbb{E}[X|\mathcal{G}]}\right] = \lim_{n \to \infty} \mathbb{E}\left[\left(\frac{1}{\mathbb{E}[X|\mathcal{G}]} \wedge n\right) \, X\right] = \lim_{n \to \infty} \mathbb{E}\left[\mathbb{E}\left[\left(\frac{1}{\mathbb{E}[X|\mathcal{G}]} \wedge n\right) \, X \Big| \mathcal{G}\right]\right]\,,
\end{equation}
where the last equality follows from the fact that the integrand is the product of a bounded random variable and an integrable random variable. Making use of the ``taking out what is known" property of conditional expectation, we have 
\begin{equation}
	\mathbb{E}\left[\mathbb{E}\left[\left(\frac{1}{\mathbb{E}[X|\mathcal{G}]} \wedge n\right) \,  X\Big| \mathcal{G}\right]\right] = \mathbb{E}\left[\left(\frac{1}{\mathbb{E}[X|\mathcal{G}]} \wedge n\right) \, \mathbb{E}\left[X | \mathcal{G}\right]\right]\,,
\end{equation}
and a further application of the monotone convergence theorem yields
\begin{equation}
	\lim_{n\to \infty}\mathbb{E}\left[\left(\frac{1}{\mathbb{E}[X|\mathcal{G}]} \wedge n\right) \, \mathbb{E}\left[X| \mathcal{G}\right]\right] = 1\,,
\end{equation}
which gives the result we set out to show.
\hfill$\Box$
\vspace{0.5cm}
\begin{prop} \label{3.3}
Let $(\pi_t)$, $(B_t)$, and $(\rho_t)$ be defined as in equations $(\ref{pk})$, $(\ref{mma})$, and $(\ref{three stars})$. If the integrability condition $\mathbb{E}[\int_0^t \pi_s\, \rd B_s] < \infty$ is satisfied for all $t \geq 0$, then $(\rho_t)_{t \geq 0}$ is a martingale.
\end{prop}
\noindent
\textbf{Proof.} It is well known that a non-negative local martingale $(x_t)_{t\geq 0}$ satisfying $\mathbb{E}[|x_0|] < \infty$ is a supermartingale, and that if 
\[
	\mathbb{E}[x_t] = \mathbb{E}[x_0]
\]	
for all $t \geq0$, then $(x_t)$ is a martingale (see e.g. Steele 2000). Our strategy will therefore be to determine a condition that ensures that $\mathbb{E}[\rho_t] = \rho_0$ for all $t \geq 0$, 
where
\begin{equation}
  \rho_0 = \mathbb{E}\left[\int_0^\infty \sigma_s^2\, \rd s\right]\,.
\end{equation}  
First we note that since $(\rho_t)$ is a supermartingale we have $\mathbb{E}[\rho_t] \leq \rho_0$, and we conclude that the random variable $\pi_tB_t$ is integrable.
Next we observe that
\begin{eqnarray}
 \mathbb{E}[\pi_tB_t] &=& \mathbb{E}\left[\mathbb{E}_t\left[\int_t^{\infty} \sigma_s^2\, \rd s\right] B_t\right] = \mathbb{E}\left[\mathbb{E}_t\left[\int_t^{\infty} \sigma_s^2\, \rd s\right] \lim_{n \to \infty}(B_t \wedge n)\right]  \\
&=&  \lim_{n \to \infty}\mathbb{E}\left[\mathbb{E}_t\left[\int_t^{\infty} \sigma_s^2\, \rd s\right] (B_t \wedge n)\right] =  \lim_{n \to \infty}\mathbb{E}\left[\mathbb{E}_t\left[\int_t^{\infty} \sigma_s^2\, \rd s \,(B_t \wedge n)\right]\right]\\
&=&  \lim_{n \to \infty}\mathbb{E}\left[\int_t^{\infty} \sigma_s^2\, \rd s\, (B_t \wedge n)\right] = \mathbb{E}\left[\int_t^{\infty} \sigma_s^2\, \rd s\, B_t\right]\,,
\end{eqnarray}
by the use of the monotone convergence theorem, the tower property, and the fact that $B_t \wedge n$ is $\mathcal{F}_t$-measurable and bounded for each $n\in \mathbb{N}$. Thus we have the identity 
\begin{equation}
	\mathbb{E}[\pi_tB_t] = \mathbb{E}\left[\exp{\left({\int_0^t \frac{\sigma_s^2}{\pi_s}\, \rd s}\right)}\,\int_t^{\infty} \sigma_s^2\, \rd s\right]\,.
\end{equation}
Our goal is to determine a condition that will ensure that the right-hand side is constant.
\noindent
To this end we observe that 
\begin{align}
& \frac{\rd}{\rd t}\left(\exp\left({\int_0^t \frac{\sigma_s^2}{\pi_s}\, \rd s}\right)\;\int_t^\infty \sigma_s^2\, \rd s \right) \nn\\
&\hspace{2cm}=  \sigma_t^2 \exp\left({\int_0^t \frac{\sigma_s^2}{\pi_s}\, \rd s}\right) \frac{\int_t^\infty \sigma_s^2\, \rd s}{\pi_t} - \sigma_t^2 \exp\left({\int_0^t \frac{\sigma_s^2}{\pi_s}\, \rd s}\right)\,.
\end{align}
Hence by the fundamental theorem of calculus, and some rearrangement, we have
\begin{align} \label{FTC}
& \exp\left({\int_0^t \frac{\sigma_s^2}{\pi_s}\, \rd s}\right)\int_t^\infty \sigma_s^2\, \rd s + \int_0^t{\sigma_s^2 \exp{\left(\int_0^s \frac{\sigma_u^2}{\pi_u}\, \rd u\right)}\, }\rd s \nn\\
&\hspace{2cm}= \int_0^t{\sigma_s^2 \exp\left({\int_0^s \frac{\sigma_u^2}{\pi_u}\, \rd u}\right)\frac{\int_s^\infty \sigma_u^2\, \rd u}{\pi_s} \, \rd s}  + \int_0^\infty{\sigma_s^2}\, \rd s\,.
\end{align}
\noindent
Now consider the first term appearing on the right-hand side of (\ref{FTC}). Since 
\begin{equation}
	\sigma_s^2 \exp\left({\int_0^s \frac{\sigma_u^2}{\pi_u}\, \rd u}\right)\frac{\int_s^\infty \sigma_u^2\, \rd u}{\pi_s} > 0 \qquad \mathbb{P}\otimes\text{Leb}-\text{a.s.}\, ,
\end{equation}
\noindent
an application of Fubini's theorem gives
\begin{equation}
	\mathbb{E}\left[\int_0^t{\sigma_s^2 \exp\left({\int_0^s \frac{\sigma_u^2}{\pi_u}\, \rd u}\right)\frac{\int_s^\infty \sigma_u^2\, \rd u}{\pi_s} \, \rd s}\right] = \int_0^t \mathbb{E}\left[\sigma_s^2 \exp\left({\int_0^s \frac{\sigma_u^2}{\pi_u}\, \rd u}\right)\frac{\int_s^\infty \sigma_u^2\, \rd u}{\pi_s}\right]\, \rd s\,.
\end{equation}
\noindent
It is shown in Lemma \ref{quotient} that $\int_s^\infty \sigma_u^2\, \rd u/\pi_s$ is integrable.
Furthermore
\begin{equation}
	\lim_{n \to \infty}\left[\sigma_s^2\exp{\left(\int_0^s \frac{\sigma_u^2}{\pi_u}\, \rd u\right)}\right]\wedge n = \sigma_s^2\exp{\left(\int_0^s \frac{\sigma_u^2}{\pi_u}\, \rd u\right)} \qquad \text{a.s.}\, .
\end{equation}
Thus, by the monotone convergence theorem, we deduce that

\begin{align}
& \int_0^t \mathbb{E}\left[\sigma_s^2 \exp\left({\int_0^s \frac{\sigma_u^2}{\pi_u}\, \rd u}\right)\frac{\int_s^\infty \sigma_u^2\, \rd u}{\pi_s}\right]\, \rd s \nn\\
&\hspace{2cm}=  \int_0^t \lim_{n \to \infty}\mathbb{E}\left[\left[\sigma_s^2\exp{\left(\int_0^s \frac{\sigma_u^2}{\pi_u}\, \rd u\right)}\right]\wedge n\,\,\frac{\int_s^\infty \sigma_u^2\, \rd u}{\pi_s}\right]\, \rd s\,.
\end{align}
\noindent
Since on the right-hand side of the equation above the integrand is given by the product of a bounded random variable and an integrable random variable, the tower property can be invoked, and we have


\begin{align} \label{interminable}
& \mathbb{E}\left[\left[\sigma_s^2\exp{\left(\int_0^s \frac{\sigma_u^2}{\pi_u}\, \rd u\right)}\right]\wedge n\,\,\frac{\int_s^\infty \sigma_u^2\, \rd u}{\pi_s}\right] \nn\\
&\hspace{2cm}=  \mathbb{E}\left[\mathbb{E}_s\left[\left[\sigma_s^2\exp{\left(\int_0^s \frac{\sigma_u^2}{\pi_u}\, \rd u\right)}\right]\wedge n\,\,\frac{\int_s^\infty \sigma_u^2\, \rd u}{\pi_s}\right]\right] \nn\\
&\hspace{2cm}=  \mathbb{E}\left[\left[\sigma_s^2\exp{\left(\int_0^s \frac{\sigma_u^2}{\pi_u}\, \rd u\right)}\right]\wedge n\,\,\mathbb{E}_s\left[\frac{\int_s^\infty \sigma_u^2\, \rd u}{\pi_s}\right]\right] \nn\\
&\hspace{2cm}= \mathbb{E}\left[\left[\sigma_s^2\exp{\left(\int_0^s \frac{\sigma_u^2}{\pi_u}\, \rd u\right)}\right]\wedge n\right]\,,
\end{align}
where the last equality follows from Lemma \ref{quotient}. More specifically, it follows as a consequence of Lemma \ref{quotient} that
\begin{equation} \label{threedots}
	\mathbb{E}_s\left[\frac{\int_s^\infty \sigma_u^2\, \rd u}{\pi_s}\right] =1\,,
\end{equation}
since the integrability of the integrand in (\ref{threedots}) entitles us to use the ``taking out what is known" property.
A further application of the monotone convergence theorem to (\ref{interminable}) then allows us to conclude that 
\begin{equation}
	\int_0^t \mathbb{E}\left[\sigma_s^2 \exp\left({\int_0^s \frac{\sigma_u^2}{\pi_u}\, \rd u}\right)\frac{\int_s^\infty \sigma_u^2\, \rd u}{\pi_s}\right]\, \rd s = \int_0^t \mathbb{E}\left[\sigma_s^2 \exp\left({\int_0^s \frac{\sigma_u^2}{\pi_u}\, \rd u}\right)\right]\, \rd s\,
\end{equation}
and by use of Fubini's theorem again, we obtain
\begin{equation} \label{fourstars}
	\mathbb{E}\left[\int_0^t{\sigma_s^2 \exp\left({\int_0^s \frac{\sigma_u^2}{\pi_u}\, \rd u}\right)\frac{\int_s^\infty \sigma_u^2\, \rd u}{\pi_s} \, \rd s}\right] = \mathbb{E}\left[\int_0^t{\sigma_s^2 \exp\left({\int_0^s \frac{\sigma_u^2}{\pi_u}\, \rd u}\right) \, \rd s}\right]  \,.
\end{equation}
On the other hand, taking expectations on both sides of equation (\ref{FTC}), we have
\begin{align} 
& \mathbb{E}\left[\exp\left({\int_0^t \frac{\sigma_s^2}{\pi_s}\, \rd s}\right)\int_t^\infty \sigma_s^2\, \rd s\right] + \mathbb{E}\left[\int_0^t{\sigma_s^2 \exp{\left(\int_0^s \frac{\sigma_u^2}{\pi_u}\, \rd u\right)}\, }\rd s\right] \nn\\
&\hspace{2cm}= \mathbb{E}\left[\int_0^t{\sigma_s^2 \exp\left({\int_0^s \frac{\sigma_u^2}{\pi_u}\, \rd u}\right)\frac{\int_s^\infty \sigma_u^2\, \rd u}{\pi_s} \, \rd s}\right]  +\mathbb{E}\left[ \int_0^\infty{\sigma_s^2}\, \rd s\right]\,.
\end{align}
Now suppose we assume that the following integrability condition is satisfied:
\begin{equation} \label{fu}
	\mathbb{E}\left[\int_0^t{\sigma_s^2 \exp{\left(\int_0^s \frac{\sigma_u^2}{\pi_u}\, \rd u\right)}\, }\rd s\right] < \infty\,.
\end{equation}
Then by use of equation (\ref{fourstars}), and writing $r_t = \sigma_t^2/\pi_t$ for brevity, we immediately deduce that
\begin{equation}
	 \mathbb{E}\left[ \exp\left({\int_0^t r_s\, \rd s}\right)\,\int_t^\infty \sigma_s^2\, \rd s \right] = \mathbb{E}\left[\int_0^\infty \sigma_s^2\, \rd s\right]\,,
\end{equation}
for all $t \geq 0$. In particular, we see that the left-hand member of the equation above is finite. Therefore we have
\begin{align} 
& \rho_0 =  \mathbb{E}\left[ \exp\left({\int_0^t r_s\, \rd s}\right)\,\int_t^\infty \sigma_s^2\, \rd s \right] =  \mathbb{E}\left[\mathbb{E}_t\left[ \exp\left({\int_0^t r_s\, \rd s}\right)\,\int_t^\infty \sigma_s^2\, \rd s \right]\right]\nn\\
&\hspace{2cm}=  \mathbb{E}\left[ \exp\left({\int_0^t r_s\, \rd s}\right)\,\mathbb{E}_t\left[\int_t^\infty \sigma_s^2\, \rd s \right]\right]  = \mathbb{E}\left[ \rho_t\right]\,,
\end{align}
which shows that $(\rho_t)$ is a martingale.
Finally, since $\rd B_t = r_tB_t\rd t$, we observe that
\begin{equation}
	\mathbb{E}\left[\int_0^t{\sigma_s^2 \exp{\left(\int_0^s r_u\, \rd u\right)}\, }\rd s\right] = \mathbb{E}\left[\int_0^t \pi_s\, \rd B_s\right]\,,
\end{equation}
and hence that (\ref{fu}) can be written in the form
\begin{equation} \label{last}
	\mathbb{E}\left[\int_0^t \pi_s\, \rd B_s\right] < \infty\,,
\end{equation}
and that concludes the proof of Proposition 3.3.
\hfill$\Box$
\vspace{0.5cm}

\begin{rem} \label{financialinterp}
The integrability condition (\ref{last}) has a financial interpretation. We consider a derivative that pays the beneficiary a continuous cash flow over the time interval $[0, t]$. The cash flow has the rate $r_sB_s$ at time $s\leq t$. In other words, the amount paid over the small interval $[s, s+ \rd s]$ is given by $\rd B_s$. The value of such a security is evidently given by $\mathbb{E}[\int_0^t \pi_s\, \rd B_s] $, and (\ref{last}) is the requirement that a finite value can be assigned to this security for any finite maturity date.
\end{rem}

\vskip 10pt \noindent {\bf Acknowledgements}.
The authors are grateful to participants at the Stockholm meeting on Recent Developments in Mathematical Finance, held in honour of Tomas Bj\"{o}rk on the occasion of his sixty-fourth birthday, for useful comments. FM would like to acknowledge financial support from the Rotary Foundation and from an EPSRC DTA grant at Imperial College London. LPH would like to acknowledge support by Shell and by Lloyds TSB.
%

\vskip 15pt \noindent {\bf References}.

\end{document}